\documentclass{webofc}
\usepackage{hyperref}
\usepackage[varg]{txfonts}   
\begin{document}

\title{Inflationary attractors from nonminimal coupling}

\author{\firstname{Zhu} \lastname{Yi}\inst{1}\fnsep\thanks{\email{yizhu92@hust.edu.cn}, Talk given by this author} \and
        \firstname{Yungui} \lastname{Gong}\inst{1}\fnsep\thanks{\email{yggong@mail.hust.edu.cn}}
}

\institute{School of Physics, Huazhong University of Science and Technology, Wuhan, Hubei 430074, China
 }

\abstract{%
 We show explicitly how the  E model attractor is obtained from the general
scalar-tensor theory of gravity with arbitrary conformal factors in the strong coupling limit.
By using conformal transformations, any
attractor with the observables $n_s$ and $r$ can be obtained. The existence of attractors imposes a challenge
to distinguish different models.
}
\maketitle
\section{Introduction}
\label{intro}
The Planck measurements of the cosmic microwave
background anisotropies give the scalar spectral tilt $n_s=0.968\pm 0.006$
and the tensor to scalar ratio  $r_{0.002} < 0.11 (95\% \text{C.L.})$\cite{Adam:2015rua,Ade:2015lrj}.
The central value of $n_s$ suggests the relation $n_s=1-2/N$ with $N=60$, here $N$ is the
number of $e$-folds when the pivotal scale exits the horizon before the end of inflation.
To the leading order of $1/N$, many inflation models have the universal result  $n_s=1-2/N$, they include the $R^2$ inflation\cite{starobinskyfr}  and the hilltop
inflation\cite{Boubekeur:2005zm}, the Higgs inflation with the nonminimal
coupling $\xi\phi^2R$\cite{Bezrukov:2007ep,Kaiser:1994vs} also gives the
same $n_s$ when $\xi\gg1$, which is a special case of the more general
nonminimal coupling $\xi f(\phi) R$ with the potential $\lambda^2 f^2(\phi)$
for an arbitrary function $f(\phi)$ in the strong coupling limit $\xi\gg1$\cite{Kallosh:2013tua}.
The above relation between $n_s$ and $N$ can be generalized to the parametrization of the observables
in terms of $N$ and used to reconstruct inflationary potential \cite{Mukhanov:2013tua,Roest:2013fha,Garcia-Bellido:2014gna,Barranco:2014ira,Boubekeur:2014xva,Chiba:2015zpa,Creminelli:2014nqa,Gobbetti:2015cya,Lin:2015fqa,Fei:2017fub,Gao:2017uja,Gao:2017owg}.

In this talk, we show that in the strong coupling limit $\xi\gg1$, any attractor is possible
for the nonminimal coupling $\xi f(\phi) R$ with an arbitrary coupling function $f(\phi)$ \cite{Yi:2016jqr}.
The E-model attractor is used as an example to support the claim. The talk is organized as follows.
In section 2, we discuss the Higgs inflation. The construction of attractor behaviours  is discussed in section 3. The conclusions are drawn in section 4.

\section{Higgs inflation}
\label{sec-1}

The simplest and easiest way to realize the cosmic inflation is to use scalar fields. Until now, the only scalar field observed is Higgs field.
The Higgs potential is
\begin{equation}\label{higgs-potentail}
  V(\phi)=\frac{\lambda}{4}\left(\phi^2-v^2\right)^2,
\end{equation}
where we set $c=\hbar=1/8\pi G=1$. Because $v\sim100\text{GeV}$ and $\phi\sim 10^{19}\text{GeV}$,
we can neglect the parameter $v$ and consider the simple power-law potential \cite{Linde:1983gd}
\begin{equation}\label{higgs-potentail2}
  V_h(\phi)=\frac{\lambda}{4}\phi^4.
\end{equation}
The action for the Higgs field minimally coupled to gravity is
\begin{equation}\label{higgs-action}
  S=\int d^4x\sqrt{-g}\left[\frac12 R-\frac12g^{\mu\nu}
\nabla_{\mu}\phi\nabla_{\nu}\phi-V_h(\phi)\right],
\end{equation}
the scalar spectral tilt $n_s$ and the tensor to scalar ratio $r$ can be calculated as
\begin{equation}\label{result-higgs1}
  n_s=0.951,\quad  r_{0.002}=0.260,
\end{equation}
with $N=60$. Obviously, this inflation model has been ruled out by the observation.

If we insist on using the Higgs field to realize the inflation, then we need to introduce the nonminimal coupling.
The action for the Higgs field nonminimally coupled to gravity is \cite{Bezrukov:2007ep,Kaiser:1994vs}
\begin{equation}\label{j-nonminimal-higgs}
  S=\int d^4x\sqrt{-\tilde{g}}\left[\frac{1+\xi\phi^2}{2} \tilde{R}(\tilde{g})-\frac12\tilde{g}^{\mu\nu}
\nabla_{\mu}\phi\nabla_{\nu}\phi-V_h(\phi)\right].
\end{equation}
By taking the  conformal transformation, in the strong coupling limit $\xi\gg1$, this action in Einstein frame becomes
\begin{equation}\label{e-nonminimal-higgs}
   S=\int d^4x\sqrt{-g}\left[\frac12 R-\frac12g^{\mu\nu}
\nabla_{\mu}\psi\nabla_{\nu}\psi-U_h(\psi)\right],
\end{equation}
where
\begin{gather}\label{e-nonminimal-higgs-psi}
 \psi=\sqrt{\frac32}\ln\left(1+\xi \phi^2\right),\\
 \label{e-nonminimal-higgs-potential}
  U_h(\psi)=\frac{\lambda}{4\xi^2}\left[1-\exp\left(-\sqrt{\frac23}\psi\right)\right]^{2},
\end{gather}
the scalar spectral tilt $n_s$ and the tensor to scalar ratio $r$ are
\begin{equation}\label{e-nonminimal-higgs-result}
  n_s\approx1-\frac2N,\quad  r\approx\frac{12}{N^2}.
\end{equation}
If we take $N=60$, then these results are consistent with the observation.

\section{The inflationary attractor}
\label{sec-2}
The action for a general scalar-tensor theory of gravity in Jordan frame is
\begin{equation}\label{general-action}
   S_J=\int d^4x\sqrt{-\tilde{g}}\left[\frac{1}{2}\Omega(\phi)\tilde{R}(\tilde{g})-\frac12\omega(\phi)\tilde{g}^{\mu\nu}
\nabla_{\mu}\phi\nabla_{\nu}\phi-V_J(\phi)\right],
\end{equation}
where $\Omega(\phi)=1+\xi f(\phi)$,   $\xi$ is a dimensionless constant, and $f(\phi)$ is an arbitrary function. Taking the conformal transformations
\begin{gather}\label{ge:gj}
  g_{\mu\nu}=\Omega(\phi){\tilde g}_{\mu\nu},\\
  \label{dpsi:dphi}
  d\psi^2=\left[\frac{3}{2}\frac{(d\Omega/d\phi)^2}{\Omega^2(\phi)}
  +\frac{\omega(\phi)}{\Omega(\phi)}\right]d\phi^2,
\end{gather}
the action \eqref{general-action} in Einstein frame becomes
\begin{equation}\label{einstein-action}
  S=\int d^4x\sqrt{-g}\left[\frac{1}{2}R(g)-\frac{1}{2}g^{\mu\nu}\nabla_\mu\psi
\nabla_\nu\psi-U(\psi)\right],
\end{equation}
where $U(\psi)=V_J(\phi)/\Omega^2(\phi)$.  From the actions \eqref{general-action}
and \eqref{einstein-action}, we see that depending on the choices of the coupling $\Omega(\phi)$,
different potentials $V_J(\phi)$ in Jordan frame become the same potential $U(\psi)$ in Einstein frame,
so we can obtain any attractor from different $V_J(\phi)$ and $\Omega(\phi)$ \cite{Yi:2016jqr}.

In reference \cite{Kallosh:2013tua}, the authors take the potential
\begin{equation}\label{universal-potential}
  V_J(\phi)=\lambda^2 f(\phi)^2=\frac{\lambda^2}{\xi^2}\left[\Omega\left(\phi\right)-1\right]^2,
\end{equation}
and   $\omega(\phi)=1$. They point out that under the strong coupling condition
\begin{equation}\label{strong-coupling}
  \frac{3(d\Omega(\phi)/d\phi)^2}{2\omega(\phi)}\gg\Omega(\phi),
\end{equation}
the scalar spectral tilt $n_s$ and the tensor to scalar ratio $r$  of this model are independent of $\Omega(\phi)$,
and the results are
\begin{gather}\label{universal-resultns}
  n_s\approx1-\frac2N,\\
  \label{universal-resultr}
  r\approx\frac{12}{N^2}.
\end{gather}
So an attractor is reached. If we set $f(\phi)=\phi^2$, this model  reduces to the  Higgs inflation model \eqref{e-nonminimal-higgs}.

The idea discussed above can be easily extended to more general cases.
For any given potential $U(\psi)$ in Einstein frame, we can obtain $n_s$ and $r$.
The same $n_s$ and $r$ can also be obtained from different potentials $V_J(\phi)$ and couplings $\Omega(\phi)$.
However, the analytical relation between $\psi$ and $\phi$ is hard to find,
so the explicit form of the potential $V_J(\phi)$ cannot be obtained. To overcome this problem,
we take the strong coupling limit for the purpose of giving analytical relation between $\psi$ and $\phi$.
The limit on $\xi$ from the strong coupling condition \eqref{strong-coupling} was discussed in \cite{Gao:2017uja}.

Under the strong coupling condition \eqref{strong-coupling}, we have
\begin{equation}\label{psi}
\psi=\sqrt{\frac{3}{2}}\ln\Omega(\phi),
\end{equation}
and   the potential in Jordan frame is
\begin{equation}\label{VJpsi}
  V_J(\phi)= \Omega^2(\phi)U\left[\sqrt{\frac{3}{2}}\ln\Omega(\phi)\right].
\end{equation}
Therefore, different potentials $V_J(\phi)$ in Jordan frame accompanied by $\Omega(\phi)$
result in the same  potential $U(\psi)$ in Einstein frame, and the same observables $n_s$ and $r$, so the attractor is obtained.

As an example, we take the E-model\cite{Kallosh:2013maa,Carrasco:2015rva}
\begin{equation}\label{Emodel}
  U(\psi)=V_0\left[1-\exp\left(-\sqrt{\frac{2}{3\alpha}}\,\psi\right)\right]^{2n},
\end{equation}
to show how to obtain the attractor. Combining Eqs. \eqref{VJpsi} and \eqref{Emodel}, we find that
\begin{equation}\label{j:Emodel:potential}
  V_J(\phi)=V_0\Omega^2(\phi)\left[1-\Omega^{-1/\sqrt{\alpha}}(\phi)\right]^{2n}.
\end{equation}
 The scalar spectral tilt $n_s$ and the tensor to scalar ratio $r$  for the potential \eqref{Emodel} are \cite{Yi:2016jqr}
 \begin{gather}\label{Emodel:ns}
  n_s=1+\frac{8 n }{3 \alpha \left[ g(N,n,\alpha)+1\right]}-\frac{8  n (n+1)}{3\alpha  \left[ g(N,n,\alpha)+1\right]^2},\\
   \label{Emodel:r}
   r=\frac{64  n^2}{3 \alpha \left[g(N,n,\alpha)+1\right]^2},
 \end{gather}
 where
 \begin{equation}\label{gfun1}
   g(N,n,\alpha)=W_{-1}\left[-\left(\frac{2n}{\sqrt{3\alpha }}+1\right) \exp \left(\frac{-4nN}{3\alpha}-\frac{2n}{\sqrt{3\alpha}}-1\right)\right],
 \end{equation}
for $n>1$ and $n/{3(2n-1)}<\alpha<{4n^2}/{3(n-1)^2}$ or $1/3<n<1$ and $\alpha>{4n^2}/{3(3n-1)^2}$;
\begin{equation}\label{gfun2}
  g(N,n,\alpha)=W_{-1}\left[-\left(\frac{2u}{3 \alpha }-\frac{2 n}{3 \alpha }+1\right)
   \exp\left({-1-\frac{2 u+2 n (2N-1)}{3 \alpha }}\right)\right],
\end{equation}\label{gfun3}
for $n>1$ and $\alpha>{4n^2}/{3(n-1)^2}$;
\begin{equation}
  g(N,n,\alpha)=W_{-1}\left[-\left(\frac{2 n}{3 \alpha }+\frac{2 v}{3 \alpha }+1\right)
 \exp\left({-1-\frac{2v+2n (2N+1)}{3 \alpha }}\right)\right],
\end{equation}
for other case; $u=\sqrt{6 \alpha  n^2+n^2-3 \alpha  n}\,$, $v=\sqrt{n (3 \alpha -6 \alpha  n+n)}
$  and  $W_{-1}$  is the lower branch of the
Lambert $W$  function.

\begin{figure}[ht]
\centering
\includegraphics[width=10cm,clip]{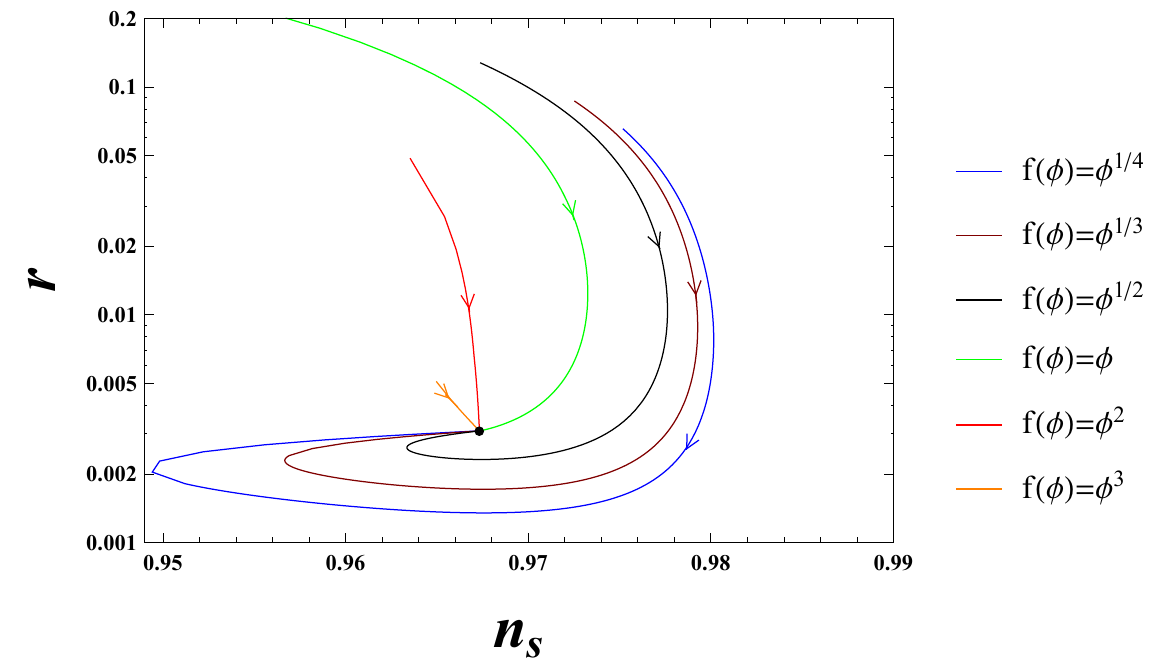}
\caption{The numerical results of $n_s$ and $r$ for the scalar-tensor theory with the potential \eqref{j:Emodel:potential}. We take
$\omega(\phi)=1$, $n=2$, $\alpha=1$, the power-law functions $f(\phi)=\phi^k$ with
$k=1/4$, 1/3, 1/2, 1, 2 and 3, and $N=60$. The coupling constant $\xi$ increases along the direction of the arrow in the plot.
The E-model attractor \eqref{Emodel:ns} and  \eqref{Emodel:r} is reached if $\xi\gtrsim 100$.}
\label{fig:Ensr}
\end{figure}

To show that the attractor \eqref{Emodel:ns} and \eqref{Emodel:r} can be
reached for arbitrary conformal factor $\Omega(\phi)=1+\xi f(\phi)$,
we take $\omega(\phi)=1$, $n=2$, $\alpha=1$, and the power-law functions $f(\phi)=\phi^k$ with
$k=1/4$, 1/3, 1/2, 1, 2 and 3.
From Eqs. \eqref{Emodel:ns},  \eqref{Emodel:r}  and  \eqref{gfun1}, we get the attractor
$n_s=0.9673$ and $r=0.0031$ for $N=60$.
We vary the coupling constant $\xi$ and choose $N=60$ to calculate $n_s$ and $r$ for
the models with the potential \eqref{j:Emodel:potential}, and the results are shown in Fig. \ref{fig:Ensr}.
From Fig. \ref{fig:Ensr}, we see that the attractor is reached in the strong coupling limit $\xi\gtrsim 100$.

\section{Conclusion}
Under the conformal transformation and the strong coupling condition \eqref{strong-coupling},
the potentials in the Jordan and Einstein frames have the relationship $V_J(\phi)=\Omega^2(\phi)U[\sqrt{3/2}\ln\Omega(\phi)]$.
For any non-singular function $\Omega(\phi)$, these models will have the same observables $n_s$ and $r$
as those given by the potential $U(\psi)$ in Einstein frame.
Therefore, these $\Omega(\phi)$ and $V_J(\phi)$ give the same attractors $n_s$ and $r$.
By this method, we may get any attractor we want, and the existence
of these $\xi$ attractors imposes a challenge to distinguish different models \cite{Yi:2016jqr}.

\textbf{Acknowledgements}:
\begin{acknowledgement}
This research was supported in part by the National Natural Science Foundation of China under Grant No. 11475065.
\end{acknowledgement}

%
%


\end{document}